\documentclass{PoS}

\title{Measurement of heavy-quark production via dielectrons in pp and p--Pb collisions with ALICE at the LHC}

\ShortTitle{Heavy-quark production via dielectrons}

\author{\speaker{Horst Sebastian Scheid}\thanks{On behalf of the ALICE Collaboration.}\\
        Goethe-University Frankfurt, Germany\\
        E-mail: \email{s.scheid@cern.ch}}

\abstract{Heavy quarks are useful probes to investigate the properties of the Quark--Gluon Plasma (QGP) produced in heavy-ion collisions at the LHC, since they are produced in initial hard scattering processes. To single out the signals that are characteristic of the QGP, it is nevertheless crucial to understand the primordial heavy-quark production in vacuum, and to disentangle hot from cold nuclear matter effects. Moreover, observations of collective effects in high-multiplicity pp and p--Pb collisions show surprising similarities with those in heavy-ion collisions. Heavy-flavour production in such collisions could give further insight into the underlying processes.

The heavy-flavour production can be studied with $\rm e^{+}e^{-}$ pairs from correlated semileptonic decays of heavy-flavour hadrons. Compared to single heavy-flavour measurements, the dielectron yield contains information about the initial kinematical correlations between the charm and anti-charm quarks, which is otherwise not accessible, and is sensitive to soft heavy-flavour production.

We report results on correlated $\rm e^{+}e^{-}$ pairs in pp collisions recorded by the ALICE detector at different collision energies. The production of heavy quarks is discussed by comparing the yield of dielectrons from heavy-flavour hadron decays as a function of invariant mass, pair transverse momentum and distance of closest approach to the primary vertex with different Monte Carlo event generators. The heavy-flavour production cross sections are also presented. Results from high-multiplicity pp collisions at $\sqrt{s} = 13$\,TeV and the status of the p--Pb analysis at $\sqrt{s_{\rm NN}} = 5.02$\,TeV are reported as well.}

\FullConference{European Physical Society Conference on High Energy Physics (EPS-HEP)\\
                10-17 July 2019\\
                Ghent, Belgium}

\begin{document}

\section{Introduction}
In relativistic nucleon--nucleon collisions dielectrons are emitted from various sources. Pseudoscalar and vector-mesons can decay directly or via Dalitz decays into a real or a virtual photon that internally converts into an $\rm e^{+}e^{-}$ pair. In addition, semi-leptonic decays from open heavy-flavour (HF) hadrons can produce a correlated dielectron when following the hadronisation and decay pattern 
\begin{equation}
    c\overline{c} \rightarrow D\overline{D} \rightarrow XY e^{+}e^{-}.
    \label{eq:cc2ee}
\end{equation}
The same holds true for dielectron pairs from open-beauty hadrons. The analysis of these pairs can then shed light on the correlation of the heavy-quark pairs, especially in the low transverse momentum regime, which is not easily accessible in other analyses.

In nucleon--nucleus collisions the before mentioned sources of dielectrons can be modified. Of particular interest are possible modifications of the heavy-flavour production via cold nuclear matter effects, e.g.\ shadowing. Additional sources of dielectrons such as thermal radiation from a hot medium, i.e.\ hadron gas or QGP, possibly formed in collisions of small systems.

\section{Data analysis}
We report on the results from two data taking periods~\cite{ref-ee7,ref-ee13}. In 2010, the ALICE detector recorded $370\times10^{6}$ minimum bias (MB) pp events at $\sqrt{s} = 7$\,TeV. In another pp data taking, in 2016, a total of $455\times10^{6}$ MB were recorded at $\sqrt{s} = 13$\,TeV. In addition, a dedicated high multiplicity (HM) trigger selected 0.036\% of the highest multiplicity pp collisions and collected $79.2\times10^{6}$ HM events.


In ALICE electrons\footnote{Electrons here and in the whole document also refers to their anti-particles, positrons.} are identified in the central barrel using the Inner Tracking System (ITS), the Time Projection Chamber (TPC), and the Time-Of-Flight system (TOF) in a kinematic range of $p_{\rm T,e} > 0.2$\,GeV/$c$ and $\eta_{\rm e} < |0.8|$. The selected electrons are then combined to opposite-sign (OS) pairs. The OS invariant mass distribution contains all correlated signal pairs, but in addition a combinatorial background. The background is estimated by constructing a spectrum of same-sign (SS) pairs. 
Residual differences in the acceptance for SS and OS pairs are estimated using event mixing and taken into account during the subtraction of the background. The spectrum is then corrected for tracking and particle identification inefficiencies.

\section{Results}

\begin{figure}[ht!]
\centering
  \begin{minipage}{0.47\textwidth}
    \includegraphics[scale=0.33,
    trim = 0 100 10 130, clip]{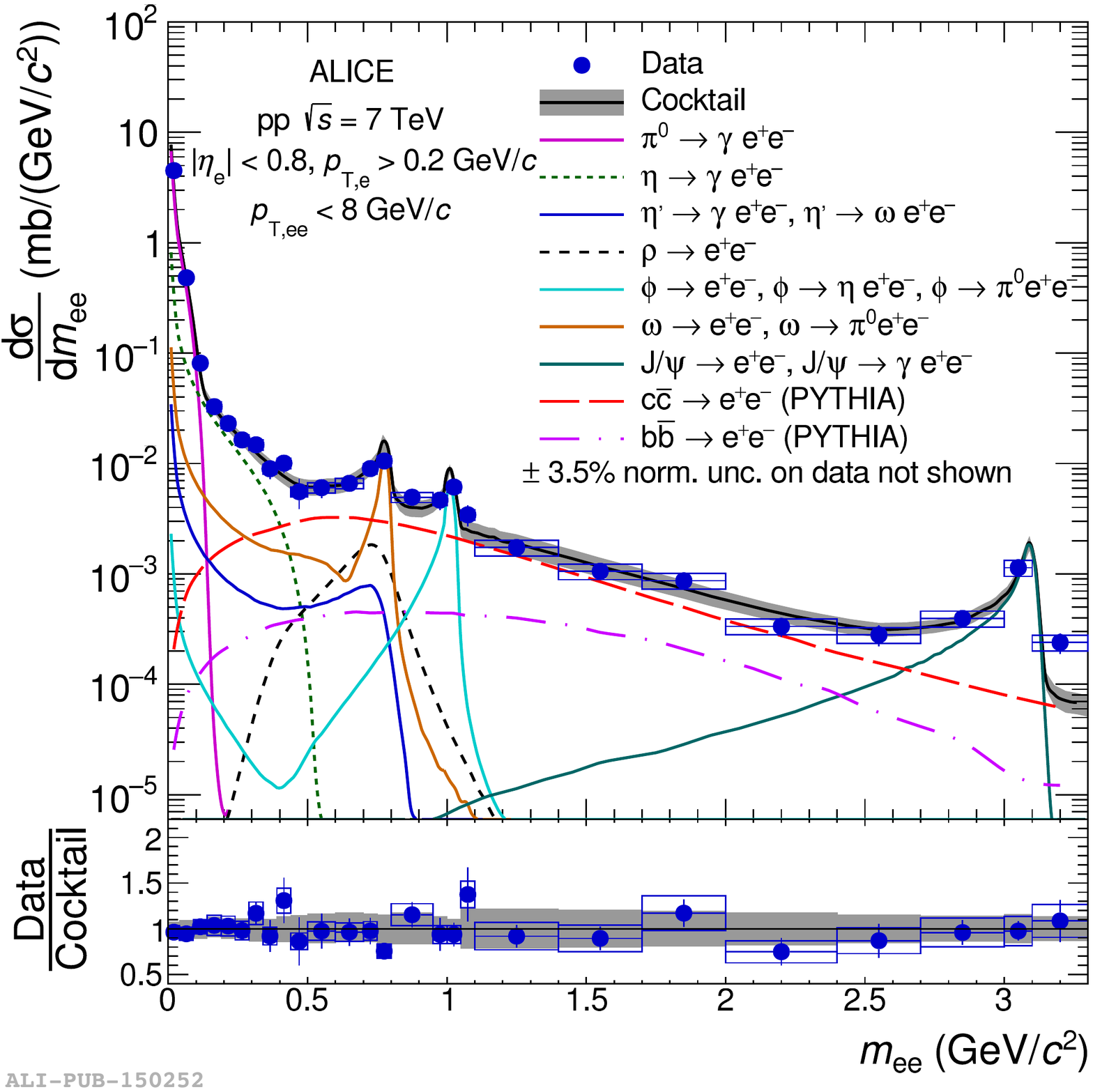}
  \end{minipage}
  \begin{minipage}{0.47\textwidth}
      \includegraphics[scale=0.33, 
      trim = 0 80 10 120, clip]{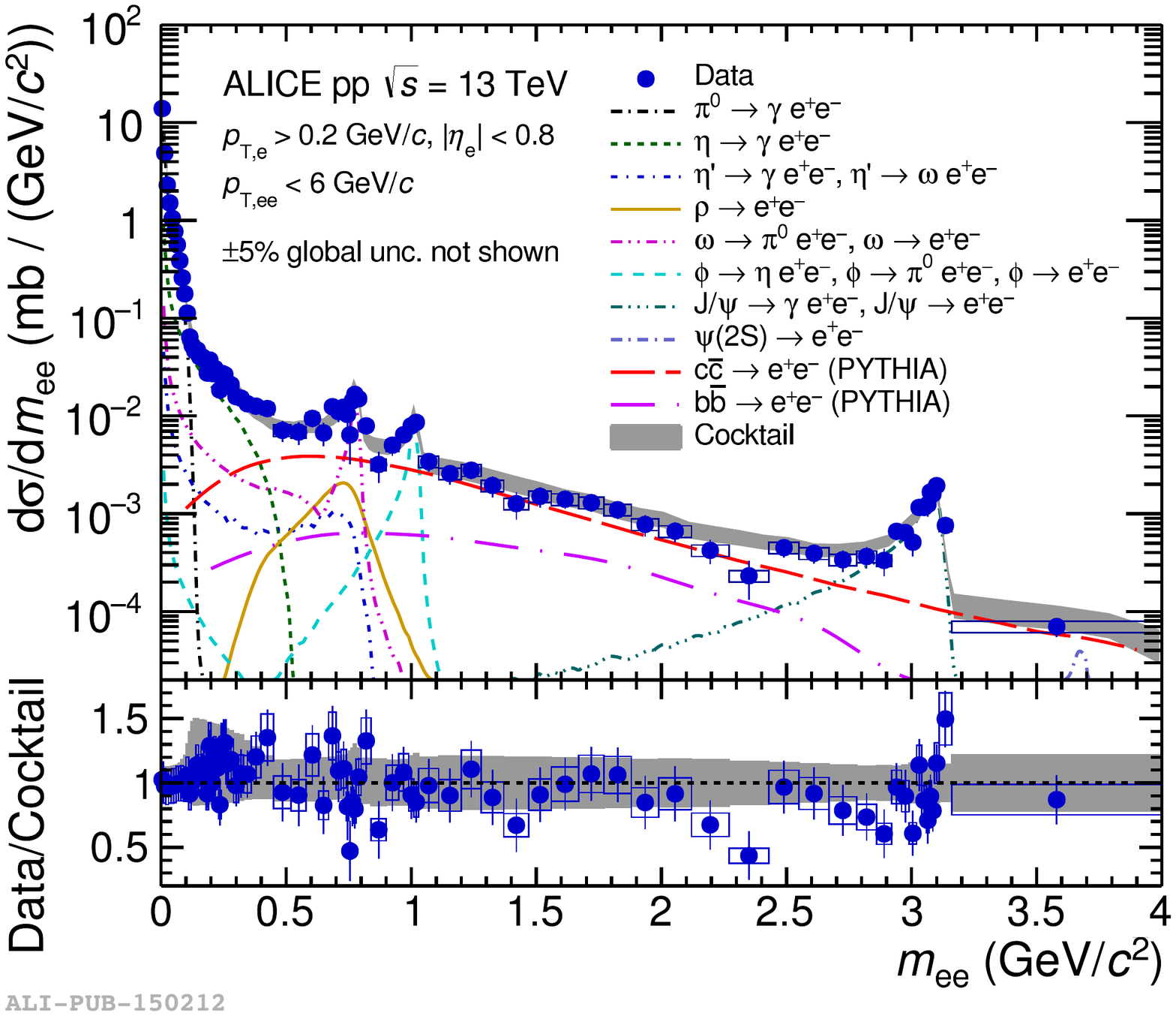}
  \end{minipage}
\caption{Dielectron cross section in pp collisions at $\sqrt{s} = 7$\,TeV (left) and $\sqrt{s} = 13$\,TeV (right) as a function of $m_{\rm ee}$ in comparison with a cocktail of known hadronic sources~\cite{ref-ee7,ref-ee13}.}
\label{fig:mee}

\end{figure}

Dielectrons were measured as a function of invariant mass ($m_{\rm ee}$) and pair transverse momentum ($p_{\rm T,ee}$) in both data taking periods. In addition, a measurement as function of $\rm DCA_{ee}$, the distance of closest approach of the electrons to the primary vertex normalised to its resolution and summed in quadrature, was performed in the 7\,TeV data sample. The $m_{\rm ee}$ spectra integrated over $p_{\rm T,ee}$, and $\rm DCA_{ee}$ are shown in Fig. \ref{fig:mee} in comparison with an expectation of the cross sections from known hadronic sources, the hadronic cocktail.

\begin{figure}[ht!]
\centering
  \begin{minipage}{0.47\textwidth}
    \includegraphics[scale=0.35,
    trim = 0 100 10 160, clip]{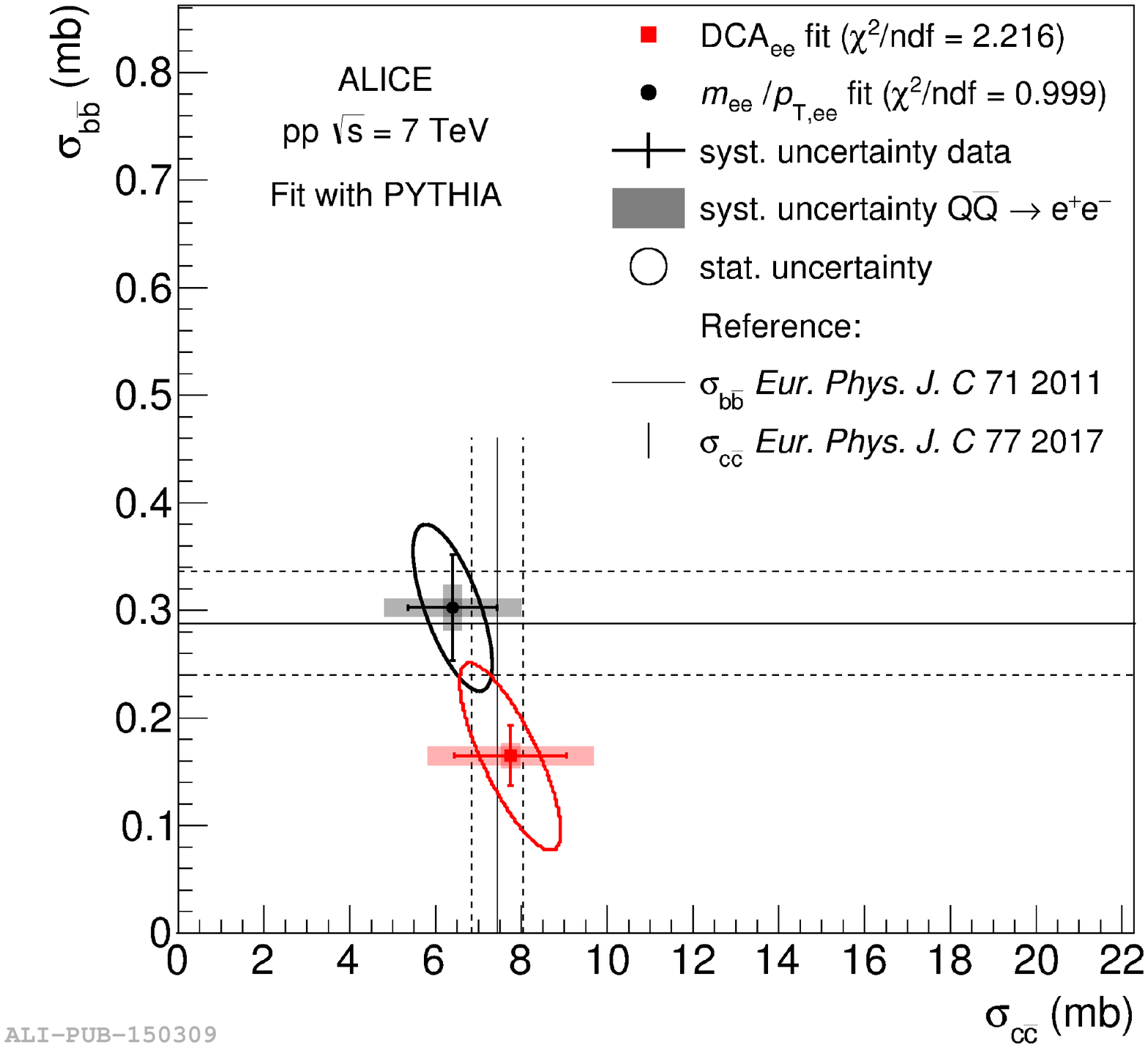}
  \end{minipage}
  \begin{minipage}{0.47\textwidth}
      \includegraphics[scale=0.35, 
      trim = 0 100 10 160, clip]{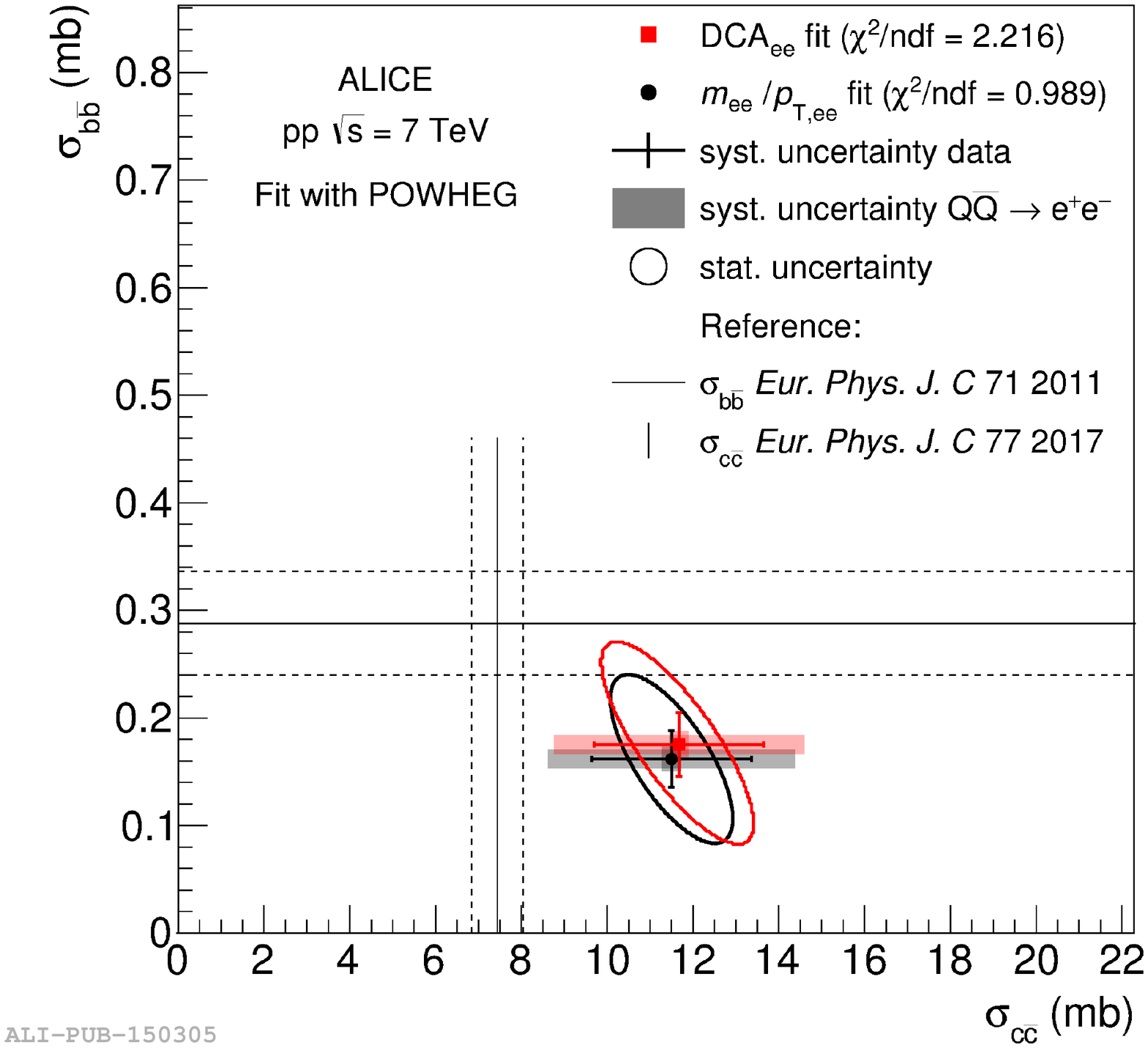}
  \end{minipage}
\caption{Total $\rm c\overline{c}$ and $\rm b\overline{b}$ cross sections with systematic and statistical uncertainties, extracted from fits of the measured dielectron yield from heavy-flavour hadron decays to ($m_{\rm ee}$, $p_{\rm T,ee}$) and to $\rm DCA_{ee}$ with PYTHIA (left) and POWHEG (right) in comparison with published cross sections from independent measurements (lines)~\cite{ref-ee7}.}
\label{fig:xsection}
\end{figure}

The measured $m_{\rm ee}$ spectra are well described by the hadronic cocktail within statistical and systematic uncertainties.
At intermediate mass both spectra are described by a contribution from charm and beauty calculated with PYTHIA6~\cite{ref-pythia6} with the Perugia2011 tune~\cite{ref-perugia2011} normalised to the cross sections measured in single heavy-flavour hadron measurements~\cite{ref-ccbar,ref-bbbar}.
In addition, it can be seen that at LHC energies the mass spectra are dominated over a wide mass range by the contribution from semi-leptonic decays of correlated open heavy-flavour hadrons.
This can be used to select and further study the production of heavy-flavour quarks in high-energy collisions. 
In the mass window of 1.1\,GeV/$c^{2}$ < $m_{\rm ee}$ < 2.7\,GeV/$c^{2}$, the so called intermediate mass region (IMR), the contributions from the HF hadrons can be selected without any significant contribution from other sources.
\begin{figure}[ht!]
\centering
  \begin{minipage}{0.47\textwidth}
    \includegraphics[scale=0.35,
    trim = 0 100 0 130, clip]{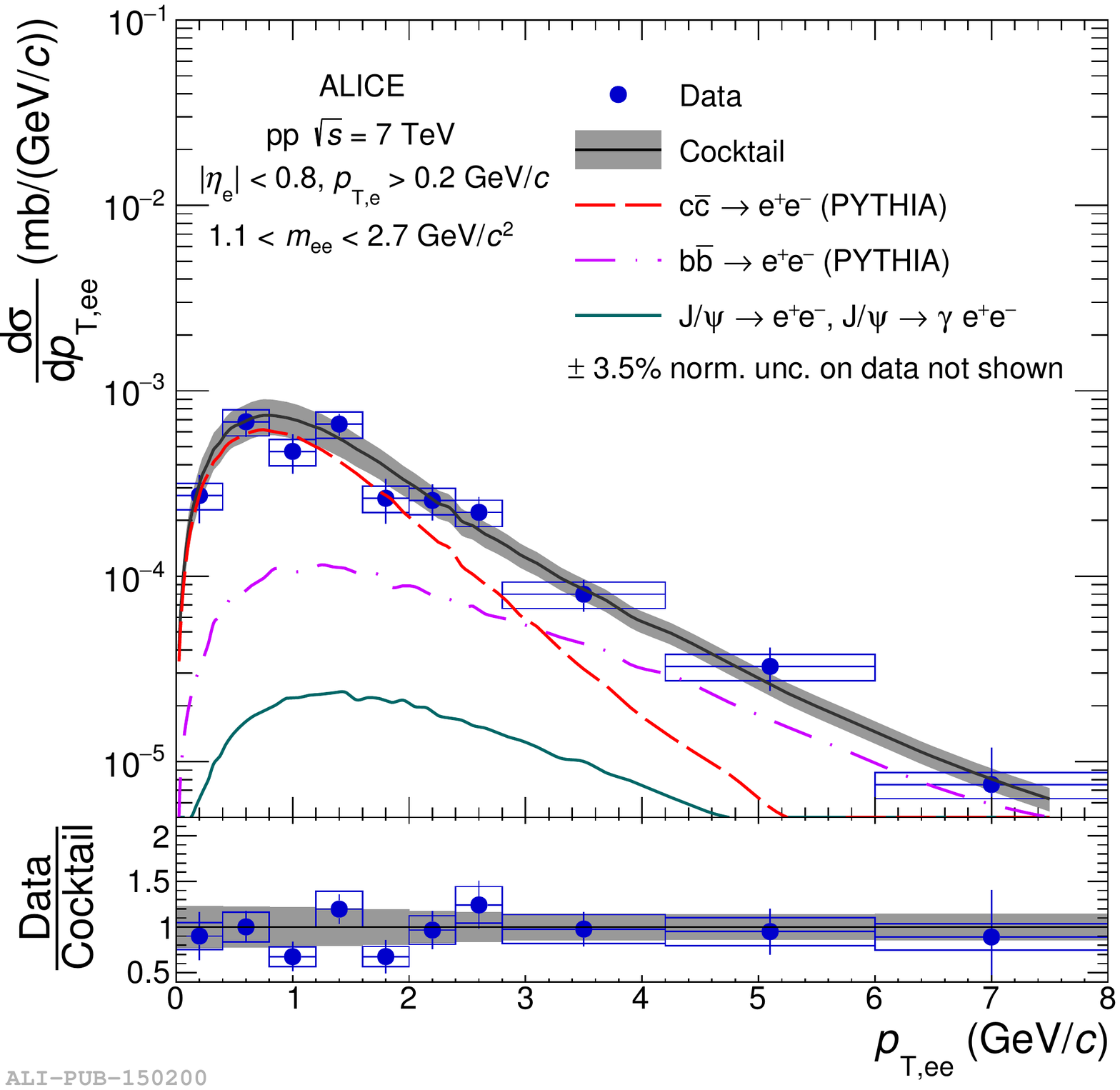}
  \end{minipage}
  \begin{minipage}{0.47\textwidth}
      \includegraphics[scale=0.35, 
      trim = 0 100 1 130, clip]{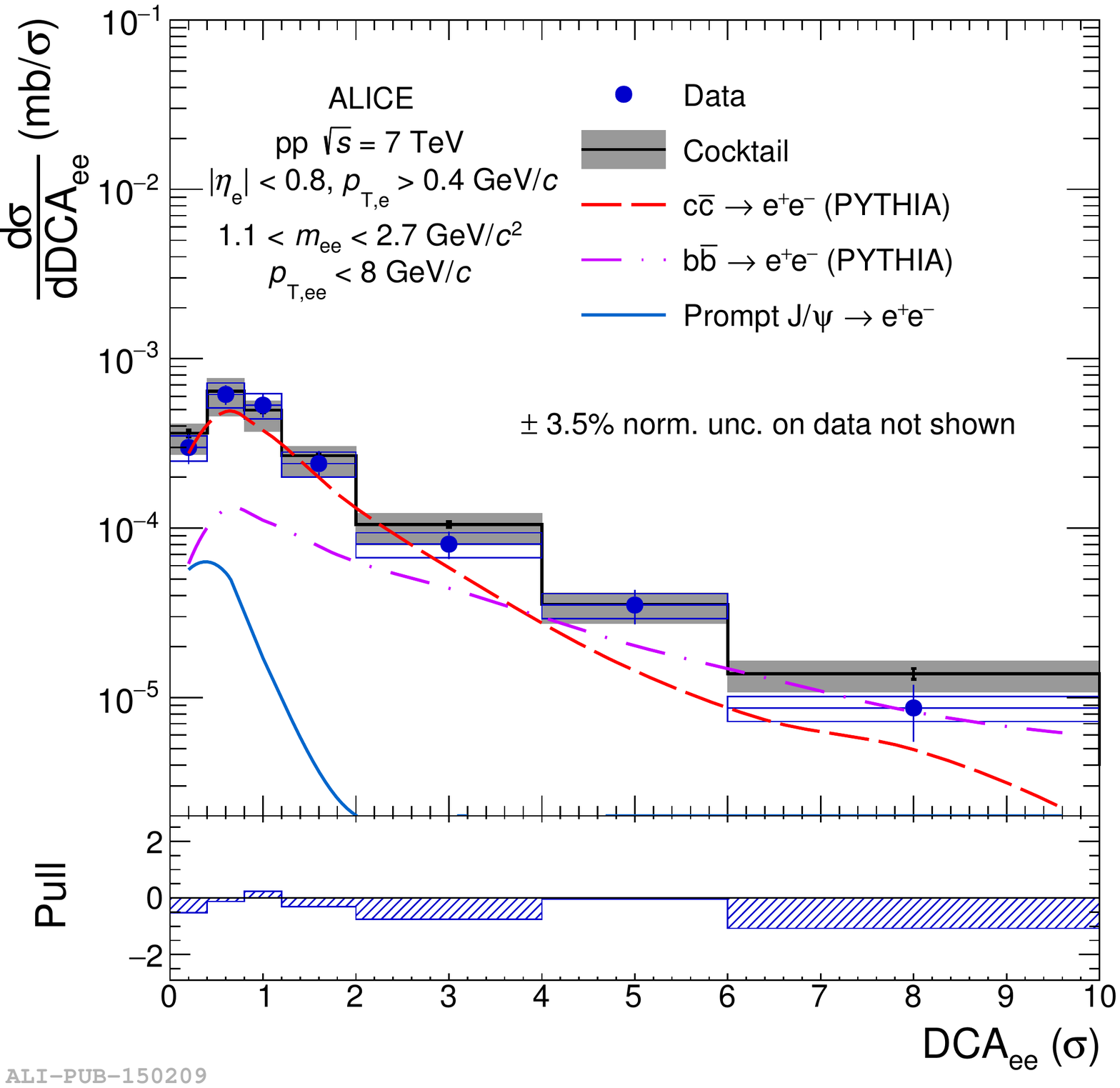}
  \end{minipage}
  
\caption{Dielectron cross section in pp collisions at $\sqrt{s} = 7$\,TeV as a function of $p_{\rm T, ee}$ (left) and $\rm DCA_{ee}$ (right) in comparison with a cocktail of known hadronic sources~\cite{ref-ee7}.}
\label{fig:pteedca}
\end{figure}
In Fig. \ref{fig:pteedca}, the $p_{\rm T,ee}$ and $\rm DCA_{ee}$ spectra measured at $\sqrt{s} = 7$\,TeV in the IMR are depicted. 
For both observables one can see that the contribution from charm and beauty have a different spectral shape. In the $p_{\rm T,ee}$ case the charm contribution dominates up to about 3\,GeV/$c$. Above this, the spectrum is dominated by the contribution from beauty quarks. For the $\rm DCA_{ee}$ observable, the crossing point is around 4$\sigma$. The distinct shapes of the two contributions in $p_{\rm T,ee}$ and $\rm DCA_{ee}$ are used to disentangle them with a two-component fit to either the $m_{\rm ee}$-$p_{\rm T,ee}$ or the $\rm DCA_{ee}$ distributions. The result is presented in Fig. \ref{fig:xsection} using heavy-flavor distributions obtained from PYTHIA, as described before, and POWHEG~\cite{ref-powheg}.
The two approaches are consistent within uncertainties, in the case of PYTHIA, as well as POWHEG. However, we can see a significant shift in the charm cross section when using POWHEG instead of PYTHIA.
This shift can be understood since PYTHIA calculates the leading-order contributions and POWHEG in addition also includes the next-to-leading order contribution, which changes the overall correlation. In a measurement of the HF cross section in the dielectron channel we are sensitive to these different contributions. The cross sections extracted with both models are in agreement with independent measurements within uncertainties.

\begin{figure}[ht]
\centering
  \begin{minipage}{0.47\textwidth}
    \includegraphics[scale=0.35,
    trim = 0 170 10 180, clip]{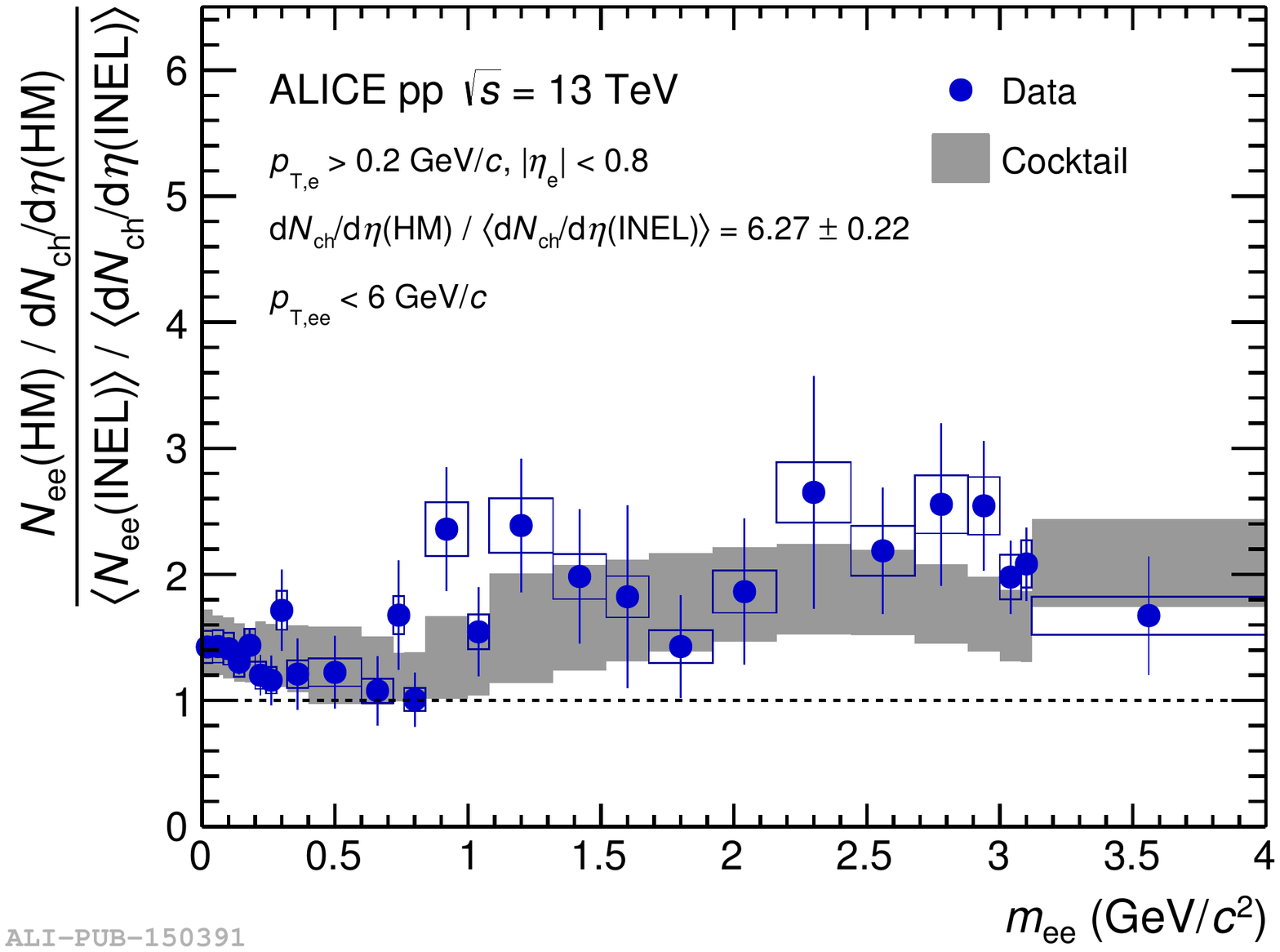}
  \end{minipage}
  \begin{minipage}{0.47\textwidth}
      \includegraphics[scale=0.35, 
      trim = 0 170 10 180, clip]{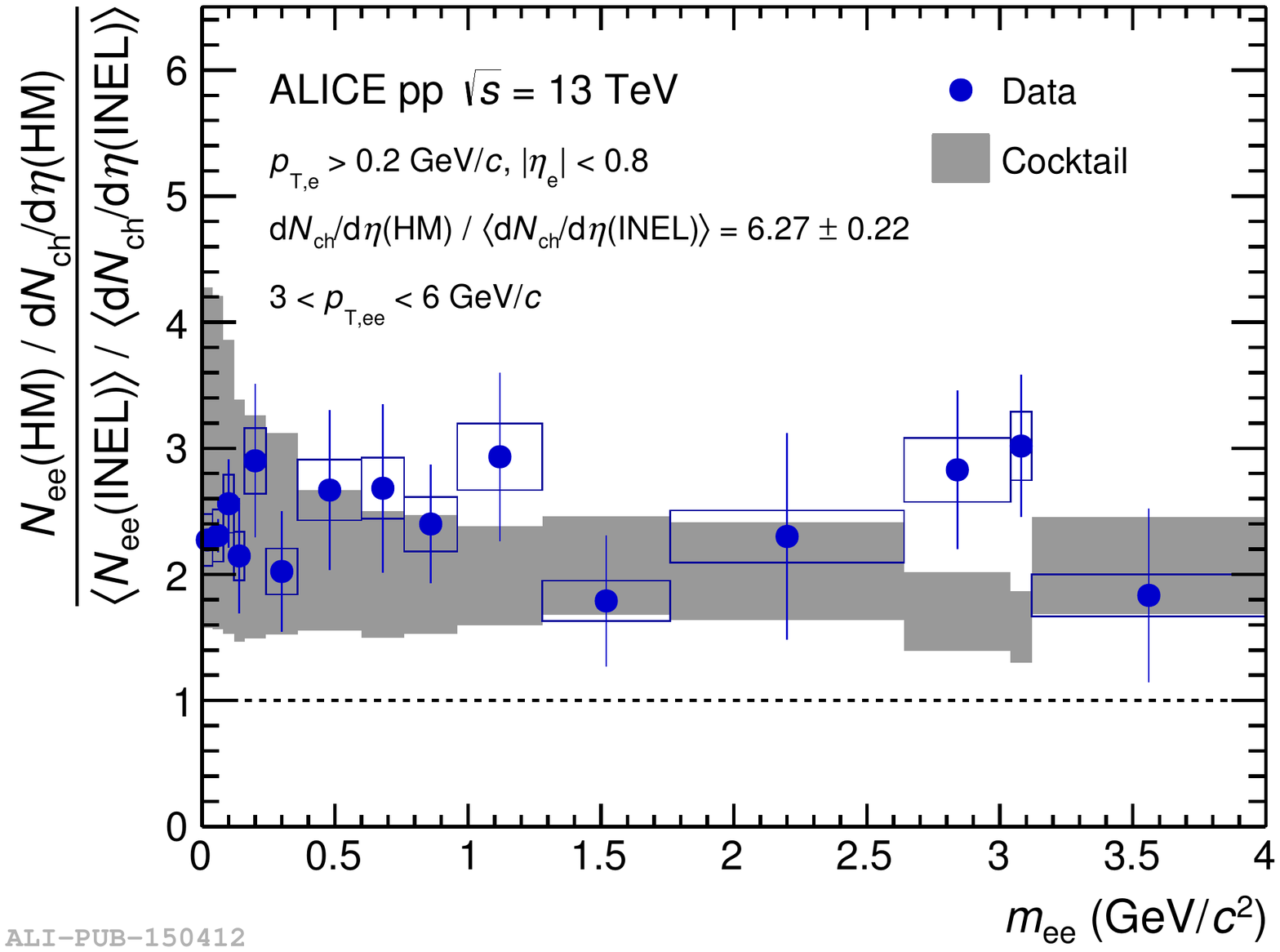}
  \end{minipage}
\caption{Ratio of dielectron production in high-multiplicity events over inelastic events integrated over $p_{\rm T,ee}$ (left) and for $p_{\rm T,ee} > 3$\,GeV/$c$ (right)~\cite{ref-ee13}.}
\label{fig:ratio}
\end{figure}
Similar findings can be reported in the 13\,TeV analysis. In this analysis the production of dielectrons was studied as function of $m_{\rm ee}$ and $p_{\rm T,ee}$ for a minimum bias and a high-multiplicity data sample.
In Fig. \ref{fig:ratio}, the ratio of the high-multiplicity dielectron spectrum over the inelastic one is shown as function of $m_{\rm ee}$, integrated over $p_{\rm T,ee}$ and for $p_{\rm T,ee} > 3$\,GeV/$c^2$, left and right, respectively. The ratios are compared with ratios of the expected hadronic contributions. The cocktail ratio reflects modifications measured independently at high multiplicity.
We use a measurement of the multiplicity dependence of D mesons~\cite{ref-DmesonsMult} to scale the heavy-flavour production, including the B mesons. For the light-flavour part of the cocktail a measurement of the multiplicity dependence of the $p_{\rm T}$ spectra is used, which shows a hardening with multiplicity~\cite{ref-ptMult}.
No significant deviation from the cocktail in both $p_{\rm T,ee}$ intervals is observed. 
The high $p_{\rm T,ee}$ part of the spectrum, dominated by the heavy-flavour contribution from beauty quarks, can be described by a cocktail constructed from D-meson measurements. This suggests a similar scaling of charm and beauty production with multiplicity at LHC energies.

In p--Pb collisions, we can further study the modification of HF hadron production resulting e.g.\ from the modification of the parton distribution functions. These modifications would be expected for small $Q^{2}$ and $x_{\rm Bj}$ in the production process of the HF quark pair and thus at low $p_{\rm T}$. This makes dielectrons a prime probe for this sort of measurements, since standard selection criteria in the analysis preserve most of the HF cross sections.
\begin{figure}[ht!]
\centering
  \begin{minipage}{0.47\textwidth}
    \includegraphics[scale=0.35,
    trim = 0 100 10 180, clip]{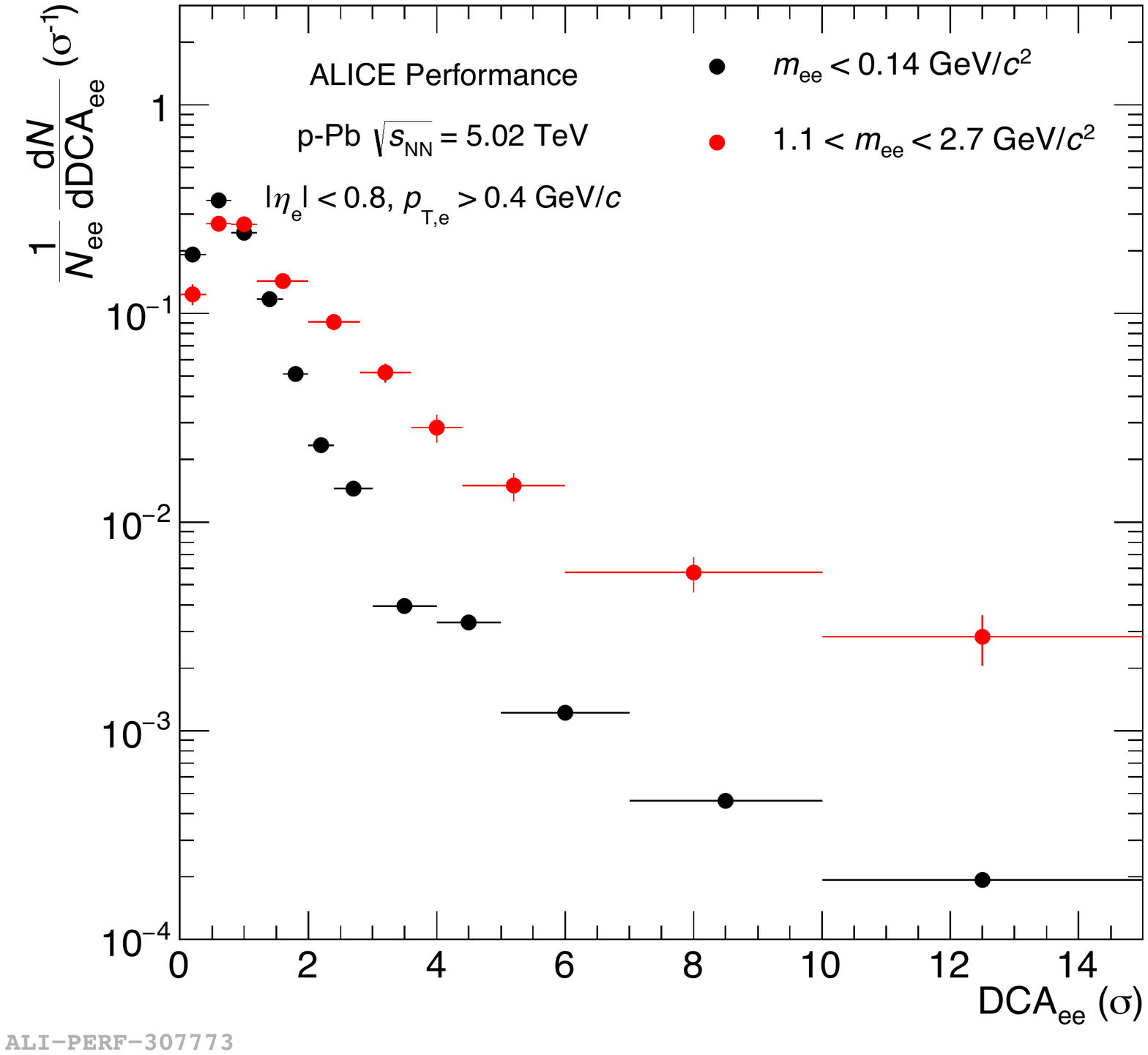}
  \end{minipage}
  \begin{minipage}{0.47\textwidth}
      \includegraphics[scale=0.35, 
      trim = 0 100 10 180, clip]{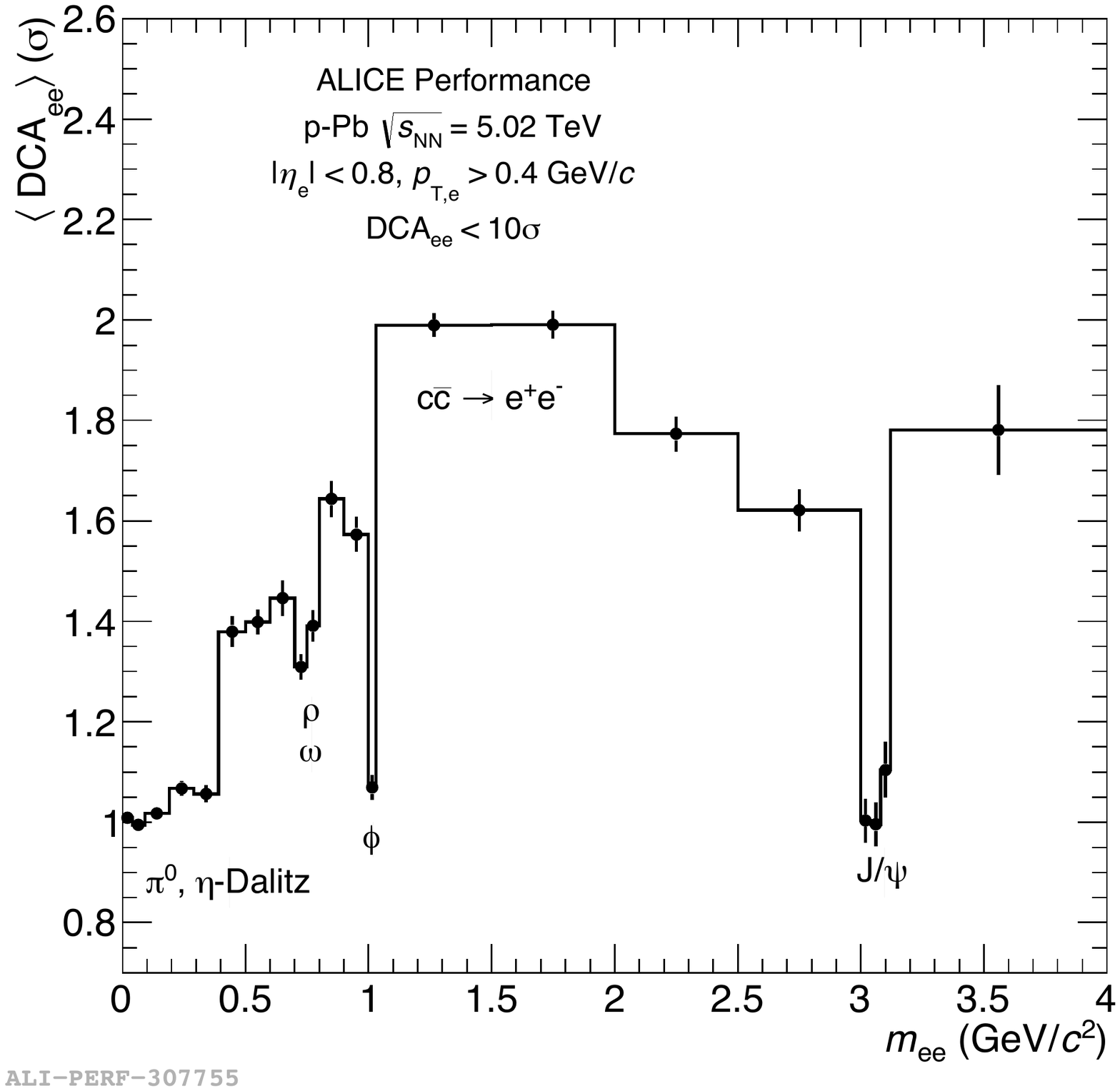}
  \end{minipage}
\caption{$\rm DCA_{ee}$ spectra in the $\pi$-mass region and the IMR normalised to unity (left) and the $\rm \langle DCA_{ee}\rangle$ as function of $m_{\rm ee}$ in p--Pb collisions at $\sqrt{s_{\rm NN}}=5.02$\,TeV.}
\label{fig:DCAppb}
\end{figure}

In Fig. \ref{fig:DCAppb} (left), we show the $\rm DCA_{ee}$ distributions for the mass region dominated by the $\pi^{0}$ Dalitz decays in comparison with the HF dominated IMR. The spectra are normalised to unity for direct comparison of the shapes. It is apparent that the HF dominated mass region has a much wider distribution. This will give the opportunity to disentangle not only the charm and beauty distributions, but in addition study a possible prompt source, such as thermal radiation. The sensitivity of the $\rm DCA_{ee}$ on the mixture of prompt and non-prompt contributions to the spectrum can be derived from Fig. \ref{fig:DCAppb} (right). We show the $\rm \langle DCA_{ee} \rangle$ as function a of $m_{\rm ee}$. For low masses ($< 0.5$\,GeV/$c^{2}$), we see a rather flat distribution. This is the region dominated by the Dalitz decays of $\pi^{0}$ and $\eta$ mesons, both prompt sources. With increasing masses the charm contribution becomes more significant, and with this the $\rm \langle DCA_{ee} \rangle$ rises. Significant drops in the distribution can be associated with the narrow contributions of the resonance decays of $\rho, \omega, \phi$ mesons. At masses larger than the $\phi$ the spectrum is completely dominated by the charm and beauty contributions and the $\rm \langle DCA_{ee} \rangle$ reaches a maximum. The falling off of $\rm \langle DCA_{ee} \rangle$ could be interpreted as a rising prompt contribution in the radiative tail of the $J/\psi$, at whose mass the spectrum is completely dominated by a prompt source again.

\section{Conclusion}

We presented the measurement of the dielectron production cross section as function of $m_{\rm ee}$, $p_{\rm T,ee}$ and $\rm DCA_{ee}$ in pp collisions at $\sqrt{s} = 7$\,TeV and as function of $m_{\rm ee}$, $p_{\rm T,ee}$ and multiplicity at $\sqrt{s} = 13$\,TeV. The spectra are well described with a hadronic cocktail for all observables. Cross sections for the production of HF quarks were extracted. The extracted cross sections are in agreement with previous measurements of single HF hadrons. A strong model dependence for this measurement points to a sensitivity to the heavy-quark production mechanisms in these models. The comparison of minimum bias and high-multiplicity $m_{\rm ee}$ spectra for $p_{\rm T,ee} > 3$\,GeV/$c$ suggests that the scaling of beauty production follows the previously observed modifications of charm production. We do not observe an indication of an additional source of thermal radiation in high multiplicity pp events within the precision of the data.
In p--Pb collisions at $\sqrt{s_{\rm NN}} = 5.02$\,TeV, the $\rm DCA_{ee}$ distribution shows promising possibilities for further studies of possible modifications of the HF contributions due to cold nuclear matter effects and to disentangle prompt and non-prompt sources. The latter would be another possibility to study possible thermal radiation in small collision systems.

\end{document}